\documentclass[10pt]{iopart}
\usepackage{graphics}
\usepackage{cite}
\begin{document}
\def\ave#1{\langle#1\rangle}
\def\Stot{S_{\rm tot}}
\def\Heff{H_{\rm eff}}
\def\Jeff{J_{\rm eff}}
\def\vS{{\bi S}}
\def\vT{{\bi T}}
\def\cH{{\cal H}}
\def\cHeff{{\cal H}_{\rm eff}}
\def\Ms{M_{\rm s}}
\def\tvT{{\tilde \vT}}
\def\hvT{{\hat \vT}}
\def\tT{{\tilde T}}
\def\hT{{\hat T}}
\jl{1}
\title
[Anomalous magnetization process in frustrated spin ladders]
{Anomalous magnetization process in frustrated spin ladders}
 \author{T\^oru Sakai\dag, Kiyomi Okamoto\ddag, Kouichi Okunishi\S\ and Masahiro Sato\ddag}

 \address{\dag Department of Physics, Tohoku University, Aoba-ku, Sendai 980-8578, Japan}

 \address{\ddag Department of Physics,
          Tokyo Institute of Technology,
          Meguro-ku, Tokyo 152-8551, Japan}
 
 \address{\S Department of Physics, Niigata University,
          Niigata 950-2181, Japan}

\begin{abstract}
We study,  at $T=0$, 
the anomalies in the magnetization curve of the $S=1$ two-leg ladder
with frustrated interactions.
We focus mainly on the existence of the $M=\Ms/2$ plateau,
where $\Ms$ is the saturation magnetization.
We reports the results by degenerate perturbation theory and
and the density matrix renormalization group,
which lead to the consistent conclusion with each other.
We also touch on the $M=\Ms/4$ and $M=(3/4)\Ms$ plateaux
and cusps.
\end{abstract}
\maketitle
%
%
\section{Introduction}
%
Anomalies in the magnetization process,
such as plateau, cusp and jump, 
in low-dimensional magnets have been attracting increasing attention
in these years.
In this paper, we investigate the effect of the frustrated interactions
on the plateaux and cusps in the magnetization curve of the $S=1$ two-leg ladder. 
Our Hamiltonian, sketched in Figure 1, is described by
\begin{eqnarray}
    \cH 
      &=&  J_0 \sum_{j=1}^N \vS_{j,1} \cdot \vS_{j,2} 
      + J_1 \sum_{j=1}^N \left(\vS_{j,1} \cdot \vS_{j+1,1} 
                          + \vS_{j,2} \cdot \vS_{j+1,2}\right) \nonumber \\
      &&+ J_2 \sum_{j=1}^N \left(\vS_{j,1} \cdot \vS_{j+1,2} 
                          + \vS_{j,2} \cdot \vS_{j+1,1}\right) \nonumber \\
      &&+ J_3 \sum_{j=1}^N \left(\vS_{j,1} \cdot \vS_{j+2,1} 
                          + \vS_{j,2} \cdot \vS_{j+2,2}\right)
                          - H \sum_{j=1}^N \left(S^z_{j,1} + S^z_{j,2}\right)
    \label{eq:Hamiltonian}
\end{eqnarray}
where $\vS_{j,l}$ is the $S=1$ operator at the $j$th site of the
$l$th ladder ($l=1,2$),
$H$ denotes the magnetic field along the $z$ direction,
and all the couplings are supposed to be antiferromagnetic
unless otherwise noticed.
\begin{figure}[ht]
   \begin{center}
      \scalebox{0.4}[0.4]{\includegraphics{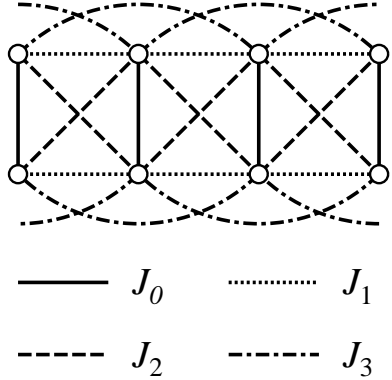}}
   \end{center}
   \caption{Sketch of the model Hamiltonian (\ref{eq:Hamiltonian}).} 
   \label{fig:model}
\end{figure}

We focus mainly on the effect of the frustrated interaction $J_2$
on the $M=\Ms/2$ plateau, 
where $M$ is the magnetization and $\Ms$ is the saturation magnetization,
although we touch on other topics.
In \S2,
we investigate the simple two-leg ladder with no frustration.
The effect of the frustrated interaction $J_2$
on the $M=\Ms/2$ plateau is discussed in \S3.
The last section \S4 is devoted to concluding remarks.

\section{$\Ms/2$ plateau of simple $S=1$ two-leg ladder}

In case of the $S=1/2$ simple ladder,
the rung interaction is always relevant.
In other words,
infinitesimally small rung interaction brings about the spin gap at $M=0$.
The situation is quite different for the $M=\Ms/2$ plateau
of the present $S=1$ model.
When $J_0 \to \infty$, the $M=\Ms/2$ plateau obviously exists,
because the problem is reduced to the two-spin problem.
On the other hand, there will be no $M=\Ms/2$ plateau
in the $J_0 \to -\infty$ limit,
because the ladder is essentially the single chain of 
$S=2$ spins formed by the rung spin pair.
Thus there exists the critical value of $J_0/J_1$.
This quantum phase transition is thought to be of the 
Berezinskii-Kosterlitz-Thouless (BKT) type \cite{Berezinskii,KT}.
We note that this $\Ms/2$ plateau state is unique (not degenerate)
from the necessary condition for the plateau \cite{OYA}.
It is interesting whether the critical point lies in the
antiferromagnetic side (i. e., $J_0>0$) or the
ferromagnetic side ($J_0<0$), when $J_1 > 0$ is fixed.
Infinitesimally small $J_0$ yields the $M=\Ms/2$ plateau if $J_0^{\rm (cr)}<0$, 
while it does not if $J_0^{\rm (cr)}>0$.

To estimate the above-mentioned critical point analytically,
we employ the degenerate perturbation theory (DPT) \cite{Totsuka}.
Hereafter we set $J_0=1$ (energy unit)
for convenience and due to the fact that
the critical point lies in the antiferromagnetic side, as will be seen later.
Let us begin with the strong rung coupling limit $J_1 \ll J_0$.
In this limit, around $M=\Ms/4$, 
we can take only two states for the rung states,
neglecting other 7 states:
the lowest state with $S_{\rm rung}^{z({\rm tot})} = 0$ and
that with $S_{\rm rung}^{z({\rm tot})} = 1$.
We can express these states by the $T^z=1/2$ and $T^z=-1/2$ states,
respectively, of the psuedo spin $\vT$.
The lowest order perturbation with respect to $J_1$ leads to
the effective Hamiltonian
\begin{eqnarray}
    &&\cHeff
    = \sum_j \left\{ 
      \Jeff^{xy} (T_j^x T_{j+1}^x + T_j^y T_{j+1}^y)
      + \Jeff^z T_j^z T_{j+1}^z 
      - \Heff T_j^z \right\} 
      \\
    &&\Jeff^{xy} = {8J_1 \over 3},~~~~ 
      \Jeff^z  = {J_1 \over 2},~~~~ 
      \Heff = H - 1 - {J_1 \over 2}.
\end{eqnarray}
The $M=0, \Ms/4$ and $\Ms/2$ states of the original $\vS$ system correspond to the
$M^{(T)}=-\Ms^{(T)},0$ and $\Ms^{(T)}$ states, respectively,
where $M^{(T)}$ ($\Ms^{(T)}$) denotes the magnetization (saturation magnetization)
of the $\vT$ system.
It is easy to obtain the field corresponding to $M^{(T)}=\Ms^{(T)}$
by considering the one-spin-down spectrum of $\cHeff$ as
\begin{equation}
   H_{\Ms/2}^{(1)}
   = 1 + {11J_1 \over 3}.
   \label{eq:lower}
\end{equation}
This $H_{\Ms/2}^{(1)}$ gives the lower edge of the $M=\Ms/2$ plateau.

Similar DPT can be developed around $M=(3/4)\Ms$, resulting in
\begin{equation}
   H_{\Ms/2}^{(2)}
   = 2 - J_1,
   \label{eq:upper}
\end{equation}
where $H_{\Ms/2}^{(2)}$ gives the upper edge of the $M=\Ms/2$ plateau.
Thus, the critical value $J_1^{\rm (cr)}$ where the $M=\Ms/2$ plateau
vanishes can be estimated from $H_{\Ms/2}^{(1)} =  H_{\Ms/2}^{(2)}$,
resulting in
\begin{equation}
    J_1^{\rm (cr)}
    = {3 \over 14}
    = 0.214.
    \label{eq:J1cDPT}
\end{equation}
We remark that there is no $M=\Ms/2$ plateau for the
so-called isotropic case $J_1 = 1$.

We have calculated the magnetization curves (Figure 2)
by use of the density matrix renormalization group (DMRG) method.
The DMRG result is consistent with the LS result $J_1^{\rm (cr)} = 0.491$,
if we consider the pathological nature of the BKT transition.
We can clearly see the $\Ms/2$ plateau $J_1=0.3$,
while we cannot for $J_1=0.5$.
It is difficult to judge for the $J_1=0.4$ case.
This situation is qualitatively consistent with the DPT results,
although the critical value is slightly larger than the
DPT prediction (\ref{eq:J1cDPT}).
This is reasonable because the plateau region will
narrowly extend like a beak of a bird on the $J_1-H$ plane,
as was seen in the diamond type spin chain case \cite{Oka-Tone}.

\begin{figure}[ht]
   \begin{center}
         \scalebox{0.28}[0.28]{\includegraphics{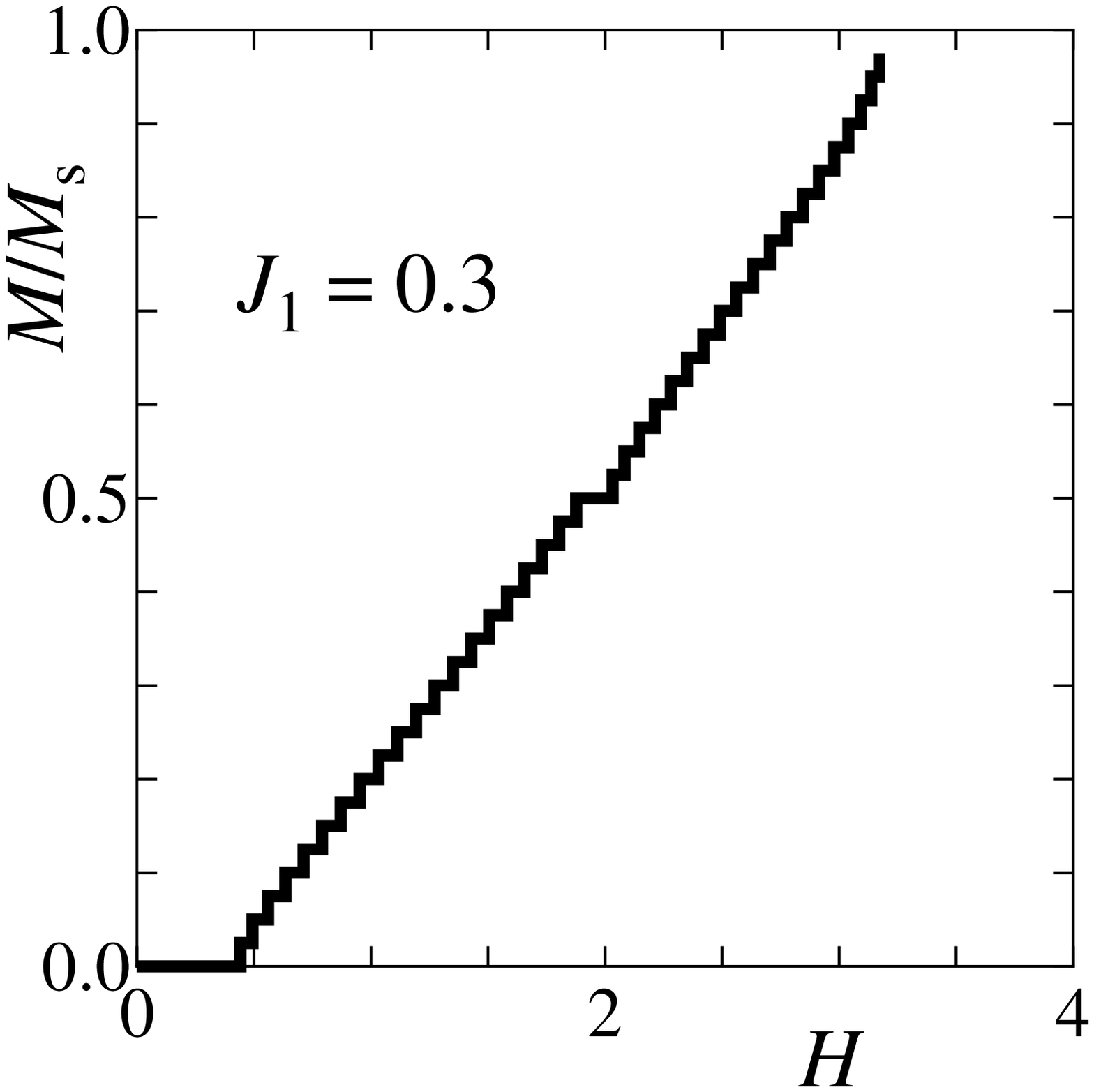}}~~~~~~~
         \scalebox{0.28}[0.28]{\includegraphics{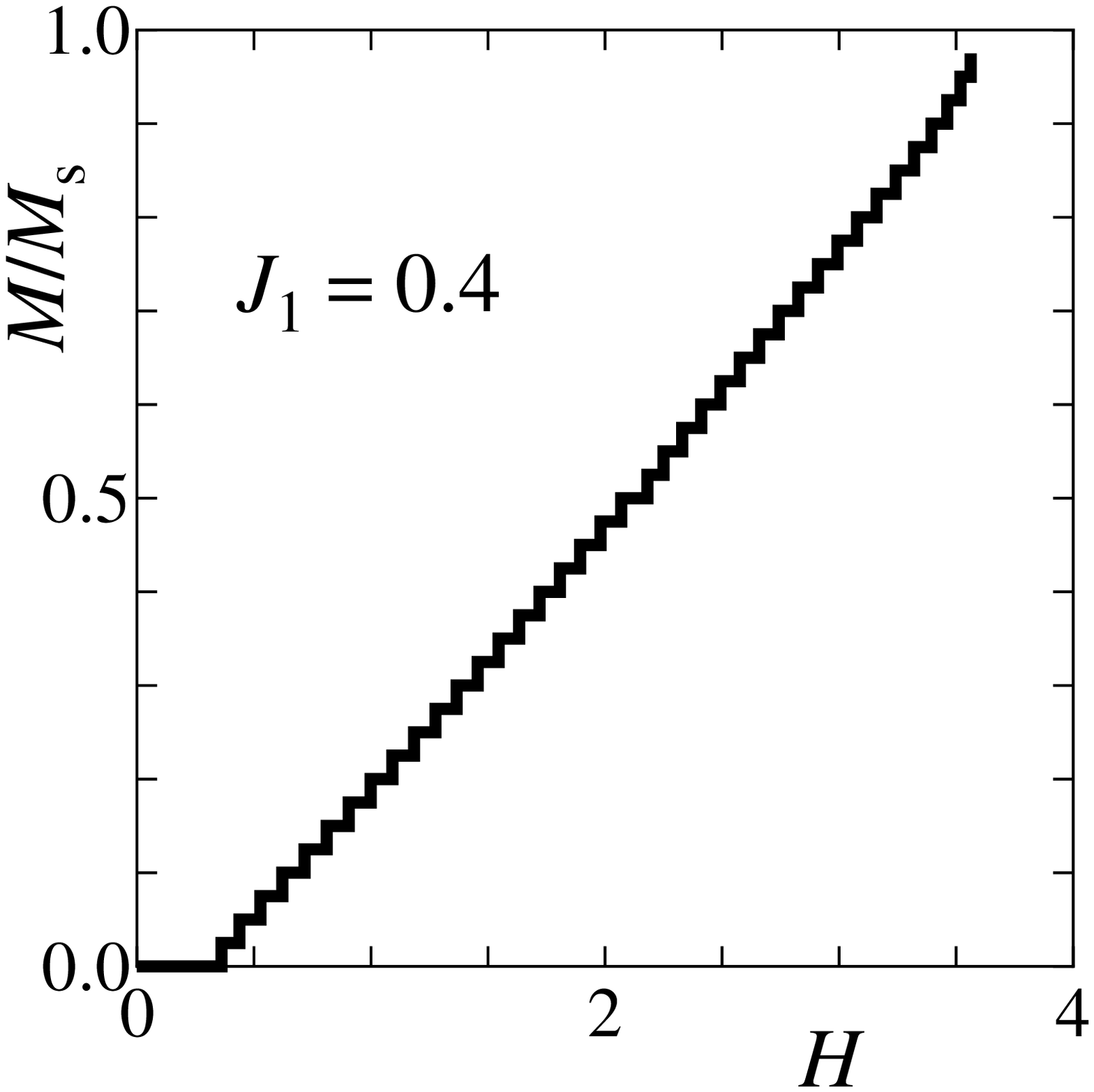}}
         \scalebox{0.28}[0.28]{\includegraphics{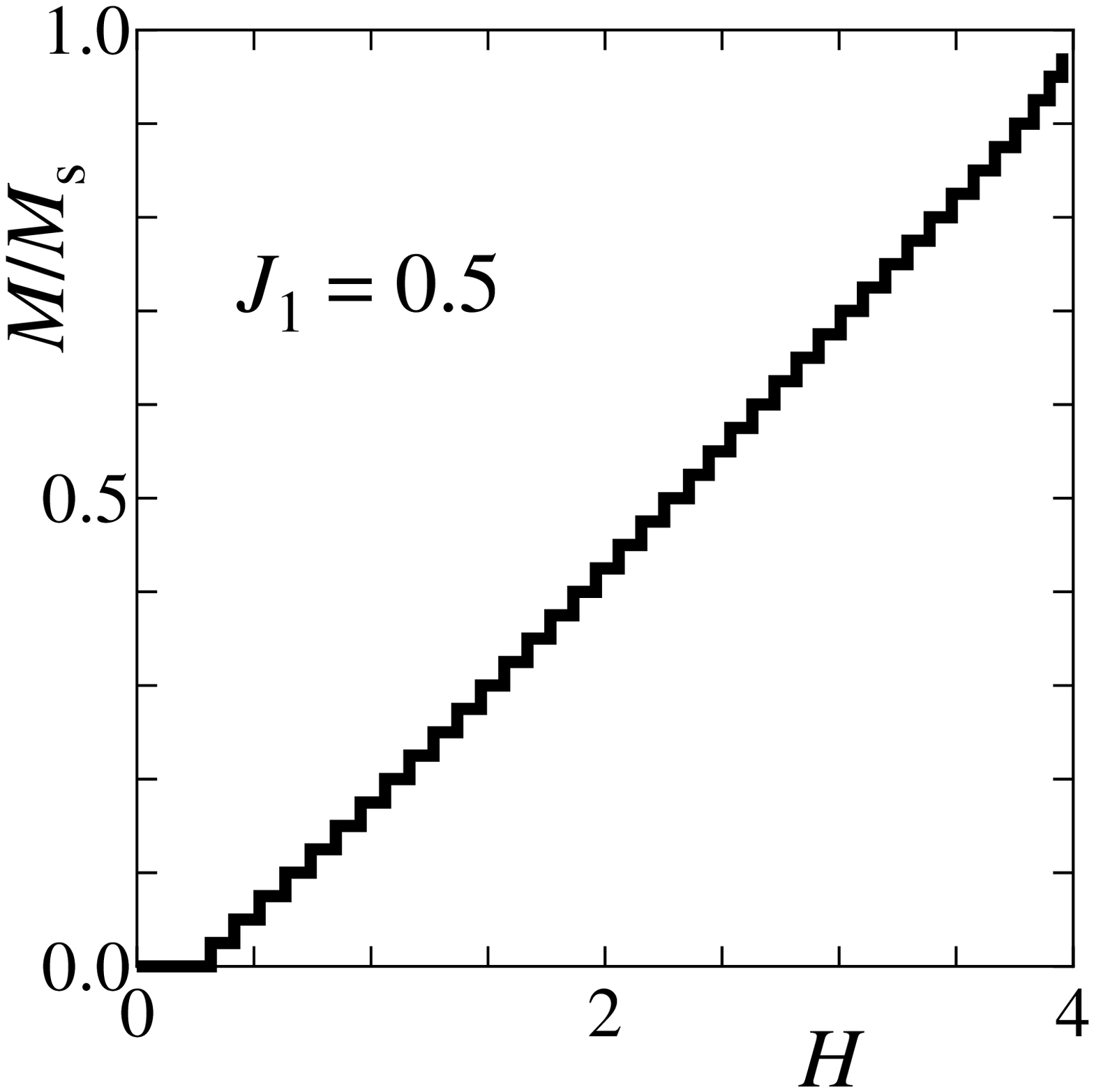}}
   \end{center}
   \caption{Magnetization curves of the simple ladder ($J_2=J_3=0$)
            when $J_1=0.3,0.4$ and $0.5$ for 40 spins obtained
            by the DMRG.
            The $\Ms/2$ plateau clearly exists for $J_1=0.3$,
            while it does not for $J_1=0.5$.
            It is difficult to judge for the $J_1=0.4$ case.}
   \label{fig:mag-curve}
\end{figure}
The level spectroscopy (LS) method \cite{Nomura-Kitazawa,Okamoto,Nomura-Kitazawa2}
is very powerful
in finding the quantum critical point of the BKT and Gaussian types.
We have also performed the LS,
the details of which will be published elsewhere.
Our conclusion is
\begin{equation}
    J_1^{\rm (cr)} = 0.491.
\end{equation}
which is consistent with the DMRG result.
We have also performed the non-Abelian bosonization approach,
finding a qualitatively consistent conclusion with those by the above methods.
Details of the LS method and the non-Abelian bosonization approach will be
published elsewhere.

\section{Effect of frustrated interactions}

Let us consider the effect of $J_2$ interactions
on the $\Ms/2$ plateau problem.
The DPT in the previous section can be easily extended to the
case with $J_2$.
We obtain
\begin{equation}
    H_{\Ms/2}^{(1)} = 1 + {11J_1 \over 3} - {5J_2 \over 3},~~~~
    H_{\Ms/2}^{(2)} = 2 - J_1 + 3J_2,
\end{equation}
yielding the critical line
\begin{equation}
    J_2 = J_1 - {3 \over 14}
\end{equation}
\begin{figure}[h]
      \begin{center}
         \scalebox{0.35}[0.35]{\includegraphics{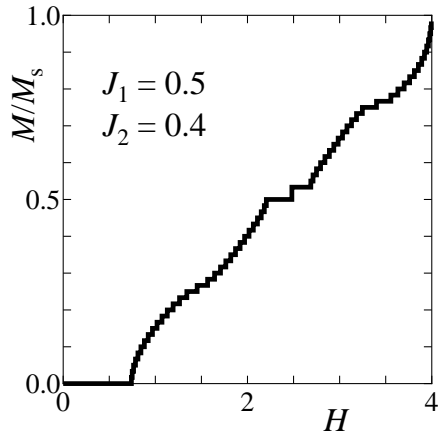}}
      \end{center}
      \caption{Magnetization curve for 60 spins when $J_1=0.5$ and $J_2=0.4$.}
      \label{fig:J1J2}
\end{figure}

The magnetization curve for the $J_1=0.5,J_2=0.4$ case 
calculated by the DMRG method is shown in Figure \ref{fig:J1J2}.
There exists the $\Ms/2$ plateau which is not seen in the 
$J_1=0.5$, $J_2=0$ case (no frustration case),
as shown in Figure \ref{fig:mag-curve}.
Furthermore we can see the $\Ms/4$ and $(3/4)\Ms$ plateaux,
which is attributed to the N\'eel ordering of the $\vT$ system,
which was first pointed out by our group \cite{oos}.
The N\'eel ordering condition is known from $\cHeff$ as
$\Jeff^{xy} < \Jeff^z$, from which we obtain
\begin{equation}
    J_2 > {13 \over 19} J_1 = 0.684J_1,
\end{equation}
for the $\Ms/4$ plateau.
This condition is satisfied for the parameter set of Figure \ref{fig:J1J2}.
The N\'eel ordering condition for the $(3/4)\Ms$ plateau is
$J_2 > 0.6J_1$,
which explains the fact that
the width of the $M=(3/4)\Ms$ plateau is wider than that of
the $\Ms/4$ plateau.
Both of the $\Ms/4$ and $(3/4)\Ms$ plateaux require the spontaneous
breaking of the translation symmetry as is known from
the necessary condition for the plateau \cite{OYA}.
We have also employed the LS for this case,
finding consistent results with that of the DMRG.

\section{Concluding remarks}
Here we shortly touch on the effect of $J_3$.
In the framework of the DPT,
the $J_3$ interaction brings about the frustrated next-nearest-neighbor
interaction between $\vT_j$ and $\vT_{j+2}$,
while the $J_2$ interaction does not.
Thus, for sufficiently large $J_3$, 
the magnetization curve has cusps as found in Figure 5.
These cusps are due to the mechanism
proposed by Okunishi, Hieida and Akutsu \cite{OHA}.
\begin{figure}[ht]
   \begin{center}
         \scalebox{0.32}[0.32]{\includegraphics{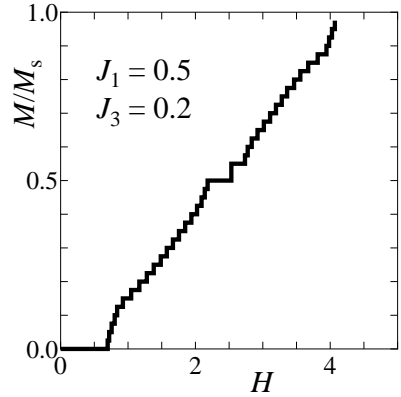}}~~~~~~~
   \end{center}
   \caption{Magnetization curve for 40 spins when $J_1=0.3$ $J_2=0$ and $J_3=0.2$.
            We can see cusps near $M=0.15\Ms$ and $M=0.85\Ms$.}
   \label{fig:cusp}
\end{figure}
The condition for the existence of the $\Ms/4$ plateau is 
$J_3 > 0.31J_1$ from the DPT \cite{oos}.
The width of the plateau may be too narrow to be observed clearly
in Figure \ref{fig:cusp}.

We have discussed the plateaux and cusps in the magnetization curve
of the $S=1$ frustrated two-leg ladder
by use of the DPT as well as the DMRG.
We have also used the non-Abelian bosonization approach and the LS method,
although we did not enter into their details.
The results obtained by these methods are consistent with each others.
We remark that the $\Ms/4$ plateau of the present mechanism
is possibly related to that observed in organic $S=1$ spin ladder 
3,3',5,5'-%
tetrakis({\it N-tert}-butylaminxyl)biphenyl 
(abbreviated as BIP-TENO) \cite{BIP}.

\end{document}